\documentclass[journal]{IEEEtran}

\usepackage{cite}
\usepackage{graphicx}
\usepackage{epstopdf}

\DeclareGraphicsExtensions{.pdf,.eps,.jpg,.jpeg,.png}

\usepackage[cmex10]{amsmath}
\usepackage{amssymb}
\usepackage{amsthm}
\usepackage{nicefrac}
\usepackage{hyperref}
\usepackage{xcolor}
\hypersetup{colorlinks=true}
\usepackage{tabularx}

\usepackage{graphicx}
\usepackage{epstopdf}
\usepackage[cmex10]{amsmath}
\usepackage{amssymb}
\usepackage{amsthm }
\usepackage{nicefrac}
\usepackage{hyperref}
\usepackage{xcolor}
\hypersetup{colorlinks=true}
\usepackage{tabularx}

\theoremstyle{theorem}

\theoremstyle{definition}

\theoremstyle{plain}

\theoremstyle{plain}

\newcommand{\sbb}{{\textbf{s}}}

\usepackage{algorithm}
\usepackage{multirow}
\usepackage{algcompatible}
\usepackage[bottom]{footmisc}

\PassOptionsToPackage{dvipsnames}{xcolor}
\usepackage{tikz}
\usetikzlibrary{calc}

\ifCLASSOPTIONcompsoc
  \usepackage[caption=false,font=normalsize,labelfont=sf,textfont=sf]{subfig}
\else
  \usepackage[caption=false,font=footnotesize]{subfig}
\fi

\usepackage{fixltx2e}
\usepackage{afterpage}
\usepackage{float}
\usepackage{stfloats}

\makeatletter

\begin{document}

\title{Trellis-Based Noise Modulation with Soft-Decision Viterbi Detection}

\author{Razieh~Torkamani, \IEEEmembership{Member,~IEEE,}
Hadi Zayyani, \IEEEmembership{Member,~IEEE,}
Mohammad Salman, \IEEEmembership{Senior Member,~IEEE,} Felipe A. P. de Figueiredo, Rausley A. A. de Souza, \IEEEmembership{Senior Member,~IEEE},

\thanks{This work was partially merged by the xGMobile Project (XGM-AFCCT-2026-8-27-1) with resources from EMBRAPII/MCTI (052/2023 PPI IoT/Manufacturing 4.0), by CNPq (302085/2025-4, 306199/2025-4), FAPEMIG (APQ-03162-24, APQ-05305-23, PPE-00124-23, RED-00194-23), by RNP Grant No. 01245.020548/2021-07, under the Brazil 6G project, and by FINEP (nº 1060/2 contract 01.25.0883.00). We also acknowledge the support of ABRINT, MODIRUM/GESPI, and NIC.br.}

\thanks{R.~Torkamani is with the Department
of Electrical Engineering, Faculty of Engineering, Bu-Ali Sina University, Hamedan, Iran (e-mail: r.torkamani@basu.ac.ir).}
\thanks{H.~Zayyani, F. A. P. de Figueiredo and R. A. A. de Souza are with the National Institute of Telecommunications (Inatel), Santa Rita do Sapucaí, Brazil. (e-mail: hadi.zayyan@posdoc.inatel.br, felipe.figueiredo@inatel.br, rausley@inatel.br).}
\thanks{M.~Salman is with the College of Engineering and Technology, American University of the Middle East, Egaila, 54200, Kuwait (e-mail: mohammad.salman@aum.edu.kw).}
}


\maketitle
\thispagestyle{plain}
\pagestyle{plain}

\begin{abstract}
Noise modulation conveys information through the statistical properties of noise rather than conventional deterministic signal parameters. This paper investigates a trellis-based noise modulation framework that exploits temporal dependencies between successive noise-power symbols. A binary filtering-based configuration is first developed as an illustrative finite-state model for comparing hard- and soft-decision sequence detection. A joint soft Viterbi receiver is then proposed, which directly incorporates the received noise energy into a likelihood-based branch metric, avoiding the information loss associated with intermediate hard decisions. The framework is further extended to an N-ary state-dependent trellis-based noise modulation scheme, where the transmitted noise-power level depends jointly on the current input symbol and trellis state. A Bhattacharyya-distance-based method is employed for systematic power-level design. Simulation results demonstrate the advantage of soft-decision sequence detection over hard-decision processing and investigate the effects of power-level design, traceback depth, energy-per-bit-to-noise ratio, symbol duration, and modulation order. The results also reveal the trade-off between spectral efficiency, detection reliability, and trellis complexity in higher-order noise modulation.
\end{abstract}


\IEEEpeerreviewmaketitle

\section{Introduction}
\label{sec:Intro}

\IEEEPARstart{N}{oise} communication is a communication paradigm in which noise is exploited as an information-bearing signal rather than being treated solely as an impairment \cite{Silva26}. In noise modulation, information is conveyed through the statistical properties of noise, such as its variance or power, instead of the amplitude or phase of a deterministic carrier. This approach can enable simple and fully non-coherent receivers without requiring oscillators, phase-locked loops (PLLs), or conventional carrier synchronization mechanisms. These characteristics make noise-based communication attractive for low-power and resource-constrained wireless systems \cite{Tashan25}, for example 6G networks \cite{Debnath26}. The concept of using thermal noise for communication can be traced back to the pioneering work of Kish \cite{Kish05}, which employed the thermal noise of resistors as information-carrying waveforms. This line of research subsequently led to several developments of Kirchhoff-Law-Johnson-Noise (KLJN) communication \cite{Kish06}-\cite{Kape22}. More recently, noise communication has been investigated from a communication-engineering perspective through noise modulation and related schemes \cite{Basar23,Basar24,Zayy26_WCL}-\cite{Khel25}.

Recent research on noise communication has explored several directions, including higher-dimensional noise modulation \cite{Zayy26_WCL}-\cite{Zayy26_CL}, extensions of conventional communication techniques to noise-based systems \cite{Zayy26_Commlett}-\cite{Anjos26}, channel estimation \cite{Shen25}-\cite{Yang26}, novel KLJN-based schemes \cite{Tasci25}-\cite{ZayyArxivRH25}, and optimal detection techniques \cite{Alshaw24}. Despite these developments, most noise modulation schemes perform detection primarily on a symbol-by-symbol basis, thereby not fully exploiting possible temporal dependencies between consecutive transmitted symbols. In conventional digital communication, sequence detection techniques such as the Viterbi algorithm exploit such dependencies through a finite-state trellis and can substantially improve detection reliability. Motivated by this observation, this work investigates the use of trellis-based sequence detection for noise modulation, where the temporal structure of the transmitted noise-power symbols is explicitly incorporated into the receiver design. Unlike a conventional Viterbi detector followed by a separate demodulation stage, the proposed soft receiver jointly exploits the statistical noise-energy information and the temporal constraints of the trellis within a unified likelihood-based detection framework.

In this paper, a finite-state noise modulation framework is investigated for both binary and higher-order signaling. The binary configuration employs a rate-one binary filtering operation to introduce a finite-state representation of the transmitted sequence and to provide an illustrative setting for comparing hard and soft likelihood-based detection. Since the binary filtering operation is invertible and does not introduce redundancy, the binary case is not intended to provide a coding gain. Rather, it serves to demonstrate the effect of preserving soft energy information before recovering the filtered sequence.

The main extension of the proposed framework is an $N$-ary state-dependent trellis-based noise modulation scheme, in which the transmitted noise-power level is jointly determined by the current state and input symbol. In contrast to the binary rate-one filtering case, this state-dependent mapping introduces nontrivial sequence constraints that can be exploited by trellis-based sequence detection.

The paper is organized as follows. Section~\ref{sec:system_model} presents the system model and problem formulation. Section~\ref{sec: prop} develops the proposed binary trellis-based noise modulation scheme and its receiver structures. Section~\ref{sec: prop1} presents the $N$-ary extension and power-level mapping. The simulation results are discussed in Section~\ref{sec: Simulation}, and conclusions are drawn in Section~\ref{sec:con}.

\section{System Model}
\label{sec:system_model}

Consider a noise-based communication system in which information is conveyed through the statistical power of a Gaussian noise sequence rather than through a conventional deterministic carrier. Let $x_t\in\{0,1\}$ denote the information symbol at time $t$, and let $\mathbf{s}_t$ denote the state of the trellis encoder. The transmitted noise samples associated with the $t$th symbol are denoted by $z_{n,t}$, $n=1,\ldots,T$, where $T$ is the number of noise samples used to represent each symbol. The samples are assumed to be independent and identically distributed according to
\begin{equation}
z_{n,t}\sim\mathcal{N}(0,P_t),
\qquad n=1,\ldots,T,
\label{eq:system_transmitted_noise}
\end{equation}
where $P_t$ denotes the noise power associated with the $t$th transmitted symbol. In the proposed framework, $P_t$ is determined by the transmitted symbol and, when a trellis structure is employed, by the current encoder state. Therefore, in its general form,
\begin{equation}
P_t=f(\mathbf{s}_t,x_t).
\label{eq:system_power_mapping}
\end{equation}

The transmitted noise sequence is observed at the receiver in the presence of additive receiver noise. Accordingly, the received samples are modeled as
\begin{equation}
y_{n,t}=z_{n,t}+w_{n,t},
\label{eq:system_received_signal}
\end{equation}
where
\begin{equation}
w_{n,t}\sim\mathcal{N}(0,\sigma_w^2)
\label{eq:system_receiver_noise}
\end{equation}
is the additive receiver noise, assumed to be independent of $z_{n,t}$. Consequently, conditioned on the transmitted symbol and trellis state, the received samples are Gaussian with zero mean and variance
\begin{equation}
y_{n,t}\sim\mathcal{N}(0,P_t'),
\qquad
P_t'=P_t+\sigma_w^2.
\label{eq:system_total_variance}
\end{equation}

Since the information is embedded in the noise power, the receiver estimates the transmitted symbol by exploiting the statistical properties of the received noise sequence. A natural sufficient statistic for detecting the noise-power level is the received energy over the $T$ samples corresponding to each symbol, defined as
\begin{equation}
E_t=\sum_{n=1}^{T}y_{n,t}^{2}.
\label{eq:system_energy}
\end{equation}
For a given power level $P_t$, the normalized energy satisfies
\begin{equation}
\frac{E_t}{P_t'}
\sim
\chi^2_T,
\label{eq:system_chi_square}
\end{equation}
which illustrates that increasing $T$ provides a more reliable estimate of the underlying received variance. This property is particularly important for noise-power-based detection and is investigated in the simulation results.

For a memoryless receiver, the decision can be obtained by comparing the observed energy with predefined thresholds corresponding to the available power levels. However, in the proposed trellis-based framework, the current symbol is not detected independently of the previous symbols. Instead, the valid transitions between the trellis states provide temporal constraints on the sequence of transmitted symbols and their associated noise-power levels. This finite-state structure can be represented by a trellis and processed using sequence-detection algorithms such as the Viterbi algorithm. The extent to which this structure provides additional detection gain depends on whether the underlying mapping introduces redundancy or nontrivial state-dependent constraints.

The proposed system considers two types of receiver processing. In the first approach, the received energy is quantized into discrete power-level decisions before trellis decoding, resulting in a hard-decision receiver. In the second approach, the unquantized received energy is directly incorporated into the likelihood-based branch metric of the Viterbi algorithm, resulting in a soft-decision receiver. The detailed binary formulation of the trellis-based noise modulator and demodulator is presented in Section~\ref{sec: prop}, while its extension to the $N$-ary case is described in Section~\ref{sec: prop1}.

\section{Binary Filtering-based Noise Modulation: An Illustrative Trellis Representation}
\label{sec: prop}
\begin{figure}[tb]
\begin{center}
\includegraphics[width=\linewidth]{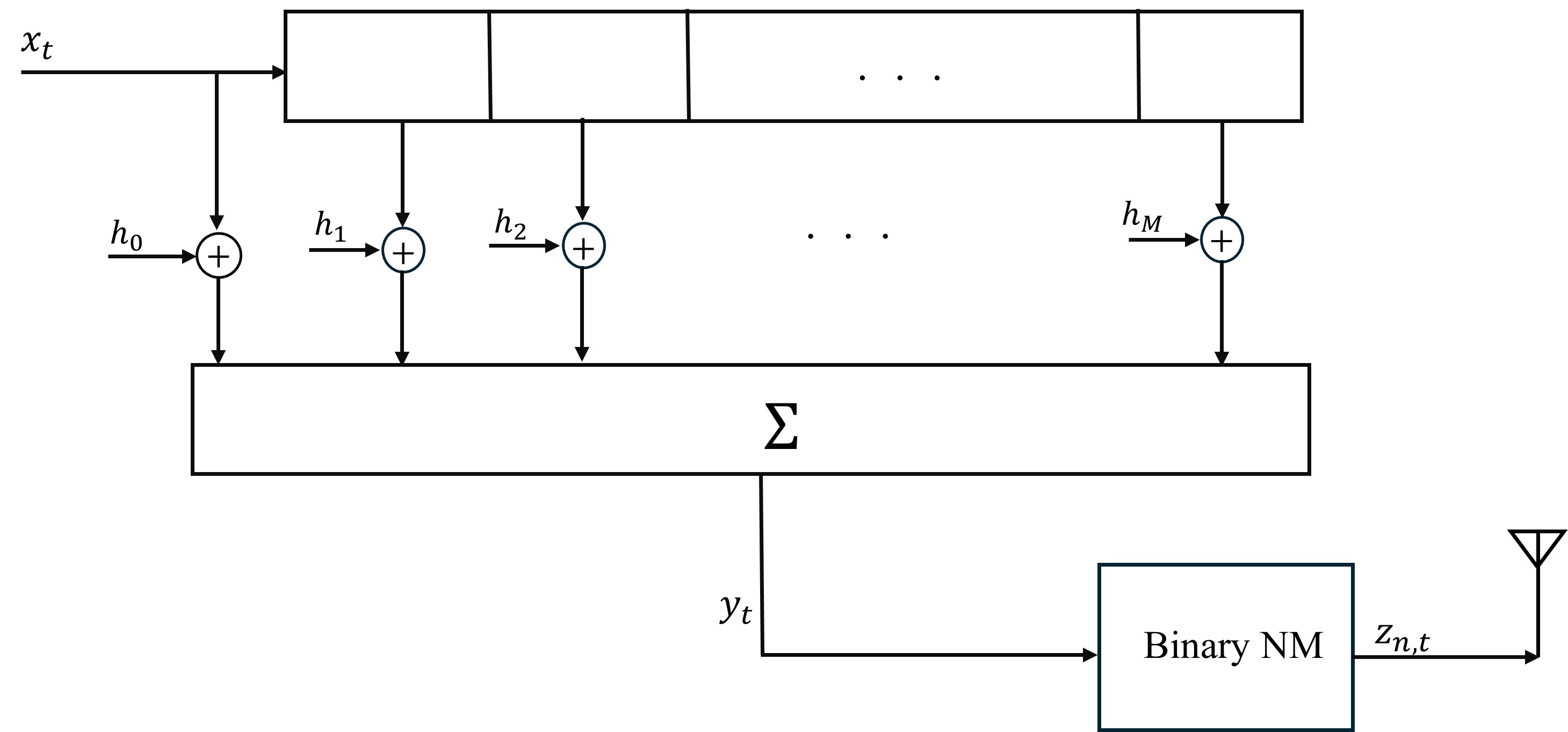}
\end{center}
\caption{Block diagram of the proposed Binary Filtering Trellis-Based Noise modulator.}
\label{fig1}
\end{figure}

In this section, a binary filtering trellis-based noise modulator-demodulator (MODEM) is proposed. The proposed structure consists of a binary filtering stage followed by a binary noise modulator, as illustrated in Fig.~\ref{fig1}. The filtering stage introduces finite memory into the binary sequence and provides a convenient trellis representation of the input-output relationship. The binary configuration considered in this section is primarily used as an illustrative case for comparing hard and soft energy-based detection and for examining the effect of finite-memory sequence recovery. 

Let $x_t\in\{0,1\}$ denote the information bit at time index $t$, where
$1\leq t\leq N_b$ and $N_b$ is the total number of transmitted bits. The output of the binary filter is denoted by $y_t\in\{0,1\}$ and is given by
\begin{equation}
y_t=
\bigoplus_{k=0}^{M} h_k x_{t-k},
\label{eq:binary_filter}
\end{equation}
where $\oplus$ denotes modulo-two addition, $h_k\in\{0,1\}$ is the $k$th binary filtering coefficient, and $M$ denotes the memory order of the filter. Thus, the filter contains $M+1$ binary coefficients. The memory introduced by the filtering operation makes the output sequence dependent on previous input bits and allows the transmission process to be represented by a finite-state trellis.

The state of the filter at time $t$ is defined as
\begin{equation}
\mathbf{s}_t=
[x_{t-1},x_{t-2},\ldots,x_{t-M}]^{T},
\label{eq:binary_state}
\end{equation}
and the corresponding next state is
\begin{equation}
\mathbf{s}_{t+1}=
[x_t,x_{t-1},\ldots,x_{t-M+1}]^{T}.
\label{eq:binary_next_state}
\end{equation}
Since each state consists of $M$ binary elements, the trellis contains
\begin{equation}
N_{\mathrm{st}}=2^M
\label{eq:number_states}
\end{equation}
states. For a given state and input bit, the filtering operation determines both the output bit $y_t$ and the next state. Therefore, each valid transition in the trellis is associated with one binary noise-modulation symbol.

The output of the binary noise modulator consists of $T$ Gaussian noise samples for each transmitted bit. Specifically,
\begin{equation}
z_{n,t}\sim\mathcal{N}(0,P_t),
\qquad n=1,\ldots,T,
\label{eq:binary_noise_modulation}
\end{equation}
where the power level $P_t$ is determined by the filtered bit $y_t$ according to
\begin{equation}
P_t=
\begin{cases}
P_H, & y_t=1,\\
P_L, & y_t=0,
\end{cases}
\qquad P_H>P_L.
\label{eq:binary_power_levels}
\end{equation}

\noindent\textit{Remark:} For the binary configuration considered here, the filtering operation in \eqref{eq:binary_filter} is rate one and causal, with $h_0=1$. Consequently, for a known initial state, the mapping between the information sequence $\{x_t\}$ and the filtered sequence $\{y_t\}$ is one-to-one. Therefore, the binary filtering stage does not introduce coding redundancy and is not expected to provide a conventional coding gain. The trellis
representation is nevertheless useful for describing the finite-memory structure and for implementing sequence recovery. In fact, maximum-likelihood sequence detection of $\{x_t\}$ is equivalent to detecting the filtered symbols $\{y_t\}$ according to their energy likelihoods and subsequently applying the inverse binary filtering operation. Hence, the binary scheme is primarily used in this work as an illustrative filtering/scrambling-based configuration for comparing hard and soft energy-based detection. The state-dependent $N$-ary scheme introduced in Section~IV constitutes the main trellis-based extension, where the power mapping depends jointly on the state and the current input symbol.

Thus, all $T$ noise samples corresponding to the same transmitted bit share the same power level. The parameters $P_H$ and $P_L$ denote the two power levels of the binary noise modulator. The particular selection of the noise-power levels is discussed later in the simulation section, where different power-level design strategies are evaluated.

\begin{figure}[tb]
\begin{center}
\includegraphics[width=0.4\linewidth]{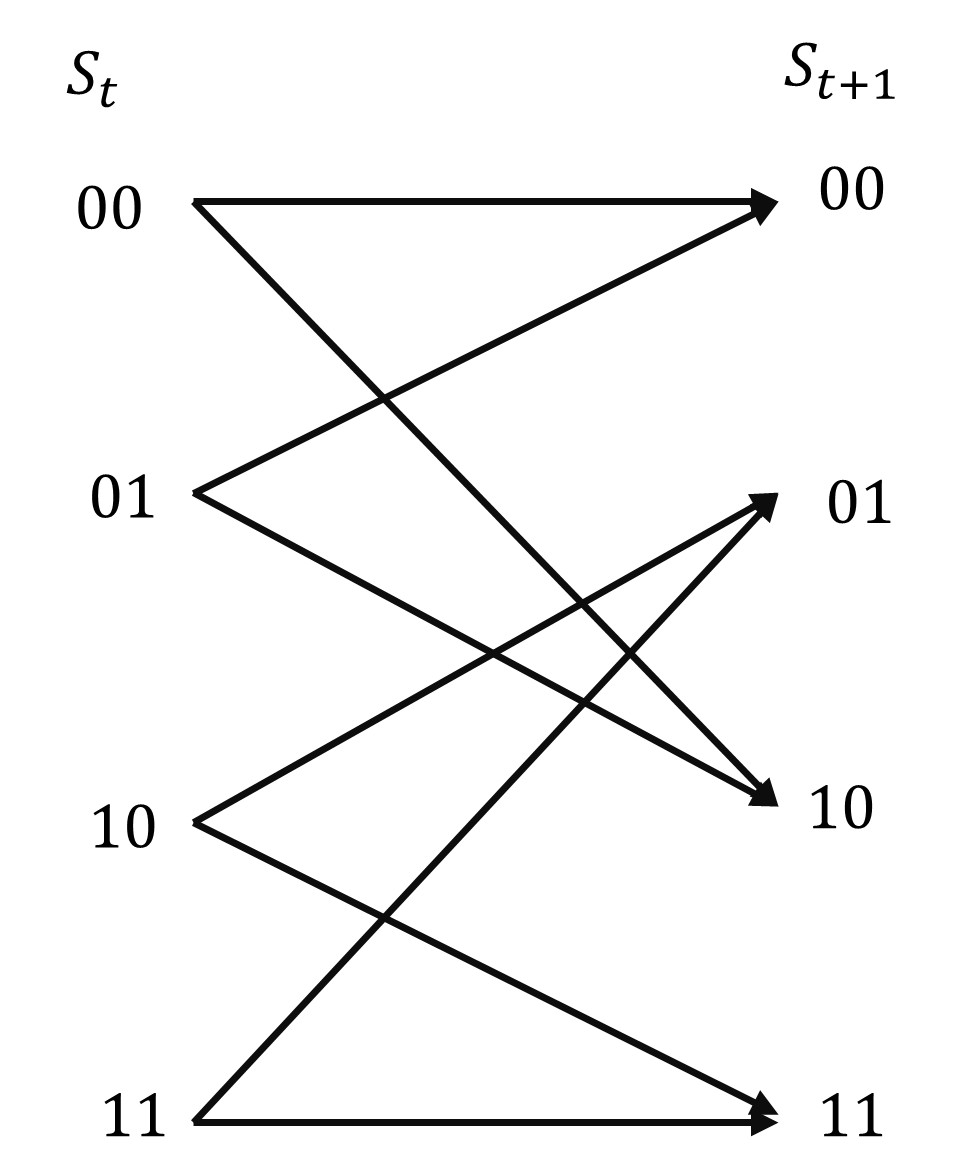}
\end{center}
\caption{The Trellis diagram of a special case of Binary filtered trellis-based noise modulator with $M=2$ and $\mathbf{h}=[h_0,h_1,h_2]^T=[1,1,1]^T$.}
\label{fig2}
\end{figure}

As an example, for $M=2$ and
\begin{equation}
\mathbf{h}=[h_0,h_1,h_2]^T=[1,1,1]^T,
\end{equation}
the corresponding trellis has $2^M=4$ states and is illustrated in Fig.~\ref{fig2}. This trellis is subsequently employed by the Viterbi algorithm for recovering the transmitted information sequence.

At the receiver, the received samples are modeled as
\begin{equation}
r_{n,t}=z_{n,t}+w_{n,t},
\label{eq:binary_received_signal}
\end{equation}
where $w_{n,t}$ represents additive white Gaussian noise with
\begin{equation}
w_{n,t}\sim\mathcal{N}(0,\sigma_w^2).
\end{equation}
Consequently, conditioned on the transmitted power level, the received samples have variance
\begin{equation}
P_t'=P_t+\sigma_w^2.
\label{eq:binary_received_variance}
\end{equation}
The received energy over the $T$ samples associated with the $t$th transmitted bit is defined as
\begin{equation}
E_t=\sum_{n=1}^{T}r^2_{n,t}.
\label{eq:binary_energy}
\end{equation}

Two receiver structures are considered for the proposed binary MODEM. In the first structure, referred to as the hard receiver, noise demodulation is performed separately from trellis sequence detection. The received energy is first converted into a binary decision, and the resulting hard decisions are then processed by a Viterbi detector. In the second structure, referred to as the soft receiver, the received energy is directly incorporated into the Viterbi branch metric, thereby avoiding the information loss associated with the intermediate hard decision. The two receiver structures are described in the following subsections.

\subsection{Separate Noise Demodulator and Viterbi Sequence Detector (Hard Receiver)}
\label{subsec:hard_receiver}

In the hard receiver, the received energy is first quantized into a
binary power-level decision and subsequently processed by the Viterbi
sequence detector. The hard decision is obtained as
\begin{equation}
\hat{y}_t=
\begin{cases}
1, & E_t\geq\mathrm{Th},\\
0, & E_t<\mathrm{Th}.
\end{cases}
\label{eq:hard_demodulator}
\end{equation}

For the two received variance levels
\begin{equation}
P_H'=P_H+\sigma_w^2,
\qquad
P_L'=P_L+\sigma_w^2,
\end{equation}
the optimum ML decision threshold is obtained by equating the
corresponding conditional likelihoods of the received energy. Since
$E_t$ follows a scaled chi-square distribution under each hypothesis,
the resulting threshold is
\begin{equation}
\mathrm{Th}_{\mathrm{ML}}
=
\frac{
T P_H'P_L'
\ln\left(P_H'/P_L'\right)
}{
P_H'-P_L'
}.
\label{eq:hard_threshold_ml}
\end{equation}

For comparison, a simpler heuristic threshold based on the midpoint
between the mean received energies is also considered:
\begin{equation}
\mathrm{Th}_{\mathrm{mid}}
=
\frac{T(P_H'+P_L')}{2}.
\label{eq:hard_threshold_mid}
\end{equation}

Unless otherwise stated, the hard receiver uses the ML threshold in
\eqref{eq:hard_threshold_ml}, while the midpoint threshold is included
in the simulation results to assess the sensitivity of the hard
receiver to the threshold selection.

The resulting hard decision $\hat{y}_t$ is then supplied to the Viterbi sequence detector. For each valid transition from $\mathbf{s}_t$ to $\mathbf{s}_{t+1}$, the branch metric is defined as the Hamming distance between the detected bit and the output bit associated with that transition:
\begin{equation}
\gamma_{\mathrm{H}}
(\mathbf{s}_t\rightarrow\mathbf{s}_{t+1})
=
d_{\mathrm{H}}
\left(
\hat{y}_t,
y_t(\mathbf{s}_t\rightarrow\mathbf{s}_{t+1})
\right),
\label{eq:hard_branch_metric}
\end{equation}
where $d_{\mathrm{H}}(\cdot,\cdot)$ denotes the Hamming distance and
$y_t(\mathbf{s}_t\rightarrow\mathbf{s}_{t+1})$ is the output bit corresponding to the considered trellis transition.

The accumulated path metric for state $\mathbf{s}_{t+1}$ is recursively calculated as
\begin{equation}
\mu_{t+1}(\mathbf{s}_{t+1})
=
\min_{\mathbf{s}_t:
\mathbf{s}_t\rightarrow\mathbf{s}_{t+1}}
\left[
\mu_t(\mathbf{s}_t)
+
\gamma_{\mathrm{H}}
(\mathbf{s}_t\rightarrow\mathbf{s}_{t+1})
\right].
\label{eq:hard_state_metric}
\end{equation}
The survivor path is retained for each state, and the transmitted information sequence is recovered through traceback. In the finite-traceback implementation, a traceback depth $D$ is employed, which determines the decoding delay.

The main limitation of this receiver is that the energy observation is converted into a binary decision before sequence detection. Consequently, information about the reliability of each energy observation is discarded. This motivates the soft-decision receiver described next, in which the complete energy observation is directly used in the trellis branch metric.

\subsection{Joint soft Viterbi detector-demodulator (soft receiver)}
\label{subsec: joint}
In the soft receiver, the energy observation $E_t$ is directly incorporated into the branch metric of the joint soft Viterbi detector-demodulator, as illustrated in Fig.~4. In this case, the branch metric is the negative of the log-likelihood of $E_t$ given the input bit. So, we will have
\begin{align}
\label{eq: bm}
\mathrm{bm}(\sbb_t\rightarrow \sbb_{t+1})&=-\ln p(E_t|x_t(\sbb_t\rightarrow \sbb_{t+1}))\nonumber\\
&=-\ln p(E_t|y_t(\sbb_t\rightarrow \sbb_{t+1})).
\end{align}
To calculate the branch metric in (\ref{eq: bm}), we note that $E_t$ is distributed as a scaled version of chi-square distribution as $E_t\sim \sigma^{'2}_t\chi^2_T$, where $\sigma^{'2}_t\triangleq\sigma^2_w+\sigma^2_t$ is equal to the variance of $r_{n,t}\sim N(0,\sigma^{'2}_t)$. So, since we know that a chi-square distributed random variable is a special case of the Gamma distribution $X\sim\chi^2_k\equiv\mathrm{Gamma}(\alpha=\frac{k}{2},\theta=2)$, we have
\begin{align}
f_X(x)=\frac{1}{2^\frac{k}{2}\Gamma(\frac{k}{2})}x^{\frac{k}{2}-1}e^{-\frac{x}{2}}.
\end{align}
Then, using $Y=AX$ with $f_Y(y)=\frac{1}{|A|}f_X(\frac{y}{A})$, we can write
\begin{align}
f_{E_t}(E_t|y_t)&=\frac{1}{\sigma^{'2}_t}f_X(\frac{E_t}{\sigma^{'2}_t})\nonumber\\
&=\frac{1}{2^\frac{T}{2}\sigma^{'2}_t\Gamma(\frac{T}{2})}\Big(\frac{E_t}{\sigma^{'2}_t}\Big)^{\frac{T}{2}-1}e^{\frac{-E_t}{2\sigma^{'2}_t}}.
\end{align}
Then, we have
\begin{align}
\label{eq: logl}
-\ln f_{E_t}(E_t|y_t)\equiv 2\ln\sigma^{'}_t-(\frac{T}{2}-1)\ln\frac{E_t}{\sigma^{'2}_t}+\frac{1}{2}\frac{E_t}{\sigma^{'2}_t},
\end{align}
where some constant terms are removed from the log-likelihood. After some other calculation on (\ref{eq: logl}) and removing some other constant terms, we arrive at the following expression for the branch metric:
\begin{align}
\mathrm{bm}(\sbb_t\rightarrow \sbb_{t+1})\equiv \gamma(\sbb_t\rightarrow \sbb_{t+1})=T\ln\sigma^{'}_t+\frac{1}{2}\frac{E_t}{\sigma^{'2}_t}.
\end{align}
The state metrics are calculated in a similar classic manner, and the same is true for the joint Viterbi detector-demodulator.

\begin{figure}[tb]
\begin{center}
\includegraphics[width=\linewidth]{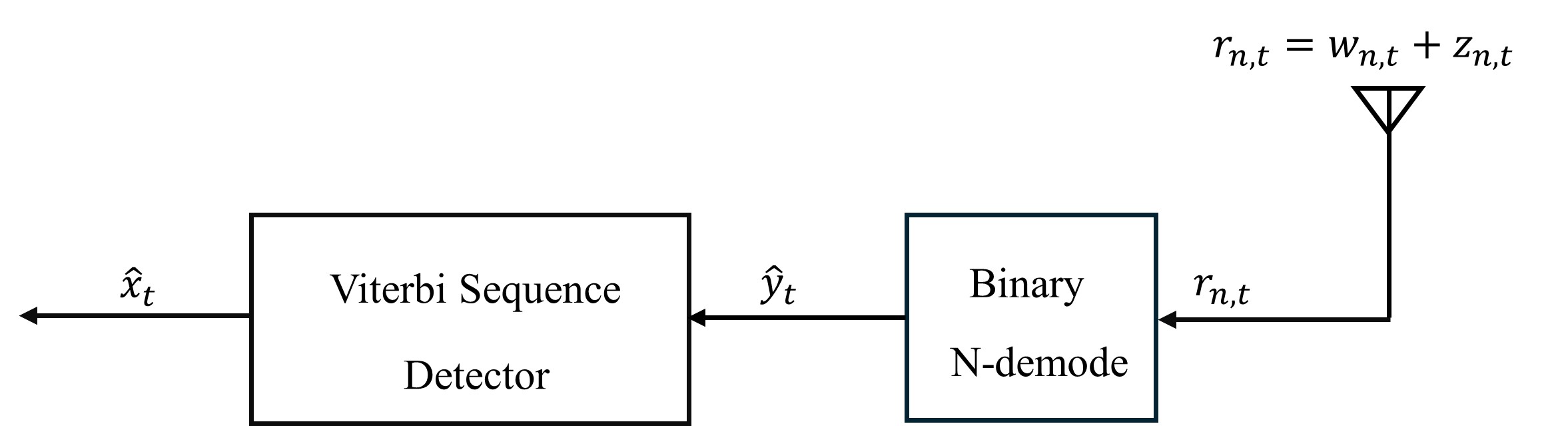}
\end{center}
\caption{Receiver structure for separate noise demodulator and hard Viterbi detector.}
\label{fig3}
\end{figure}

\begin{figure}[tb]
\begin{center}
\includegraphics[width=\linewidth]{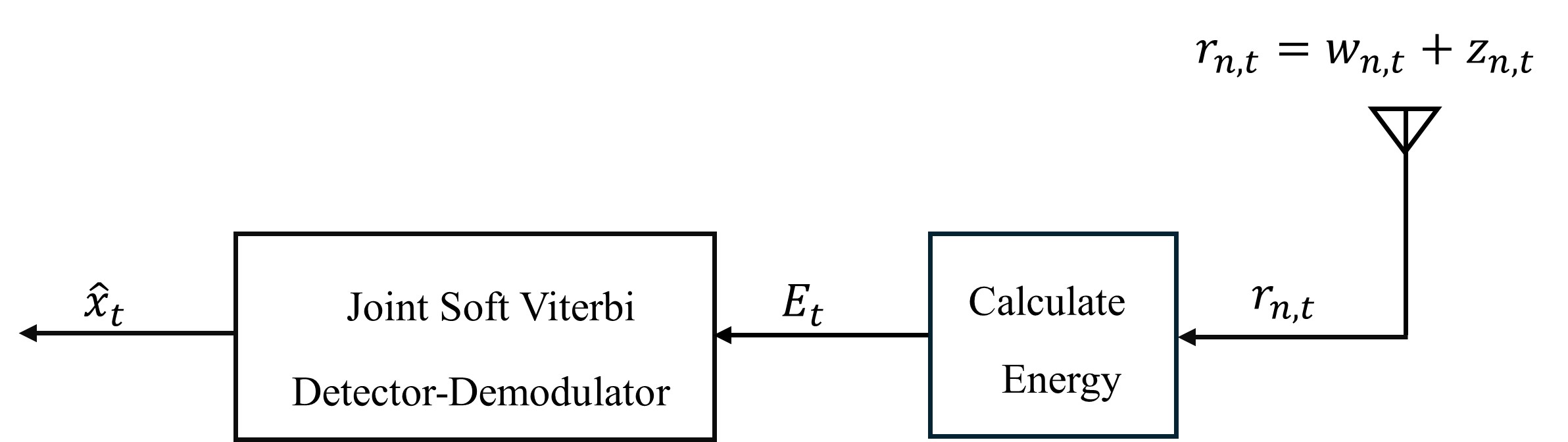}
\end{center}
\vspace{+0.5 cm}
\caption{Receiver structure for joint Viterbi-based detector-demodulator.}
\label{fig4}
\end{figure}

\begin{figure}[tb]
\begin{center}
\includegraphics[width=\linewidth]{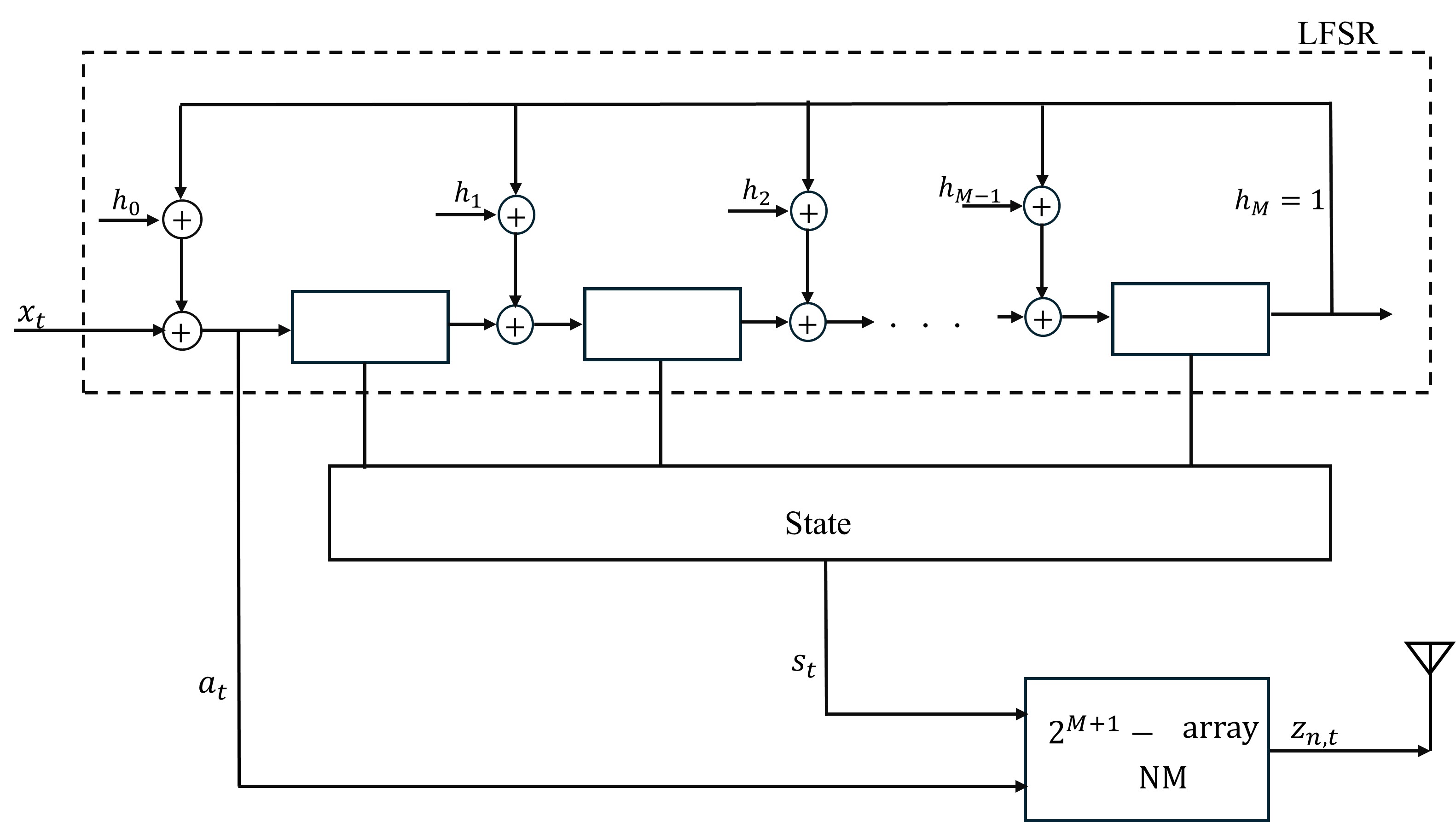}
\end{center}
\caption{The proposed N-ary Trellis-based noise modulator.}
\label{fig5}
\end{figure}
\section{Proposed $N$-ary Trellis-Based Noise Modulation and Demodulation}
\label{sec: prop1}

The binary configuration of Section~\ref{sec: prop} provides a simple finite-memory and invertible filtering example for illustrating the difference between hard and soft energy processing. Since the binary filter is rate one, it does not introduce conventional coding redundancy. In this section, the framework is generalized to an $N$-ary state-dependent trellis-based noise modulation scheme, where the modulation order and the trellis memory are explicitly defined by separate parameters.

Let the transmitted symbol at time index $t$ be denoted by
\begin{equation}
u_t\in\mathcal{A}_N,
\qquad
\mathcal{A}_N={0,1,\ldots,N-1},
\label{eq:nary_alphabet}
\end{equation}
where $N$ is the modulation order, i.e., the number of possible input symbols per transmitted symbol interval. Thus, each input symbol carries $\log_2N$ bits when a binary representation of the alphabet is used.

The trellis memory order is denoted by $M$. The state at time $t$ is defined by the previous $M$ input symbols as
\begin{equation}
\mathbf{s}_t=
[u_{t-1},u_{t-2},\ldots,u_{t-M}]^T.
\label{eq:nary_state}
\end{equation}

Since each memory element can take one of $N$ possible values, the total number of trellis states is
\begin{equation}
N_{\mathrm{st}}=N^M.
\label{eq:number_states}
\end{equation}

For a transition associated with the current input symbol $u_t$, the next state is
\begin{equation}
\mathbf{s}_{t+1}=[u_t,u_{t-1},\ldots,u_{t-M+1}]^T.
\label{eq:nary_next_state}
\end{equation}

Accordingly, each state has $N$ outgoing branches, corresponding to the $N$ possible values of $u_t$. The total number of trellis transitions is therefore
\begin{equation}
N_{\mathrm{tr}}=N^{M+1}.
\label{eq:number_transitions}
\end{equation}

For each trellis transition, the transmitted noise-power level is determined by a state-dependent mapping
\begin{equation}
P_t=f(\mathbf{s}_t,u_t),
\label{eq:power_mapping}
\end{equation}
where $f(\cdot)$ maps a state-input pair to one of the available noise-power levels. Importantly, the number of available power levels is not required to be equal to the modulation order $N$ or to the number of trellis transitions. Let $Q$ denote the number of distinct noise-power levels used by the modulation scheme, with
\begin{equation}
1\leq Q\leq N^{M+1}.
\label{eq:number_power_levels}
\end{equation}

The mapping in \eqref{eq:power_mapping} can therefore be one-to-one when $Q=N^{M+1}$, or multiple trellis transitions may share the same power level when $Q<N^{M+1}$. This distinction allows the modulation order $N$, the memory order $M$, and the number of noise-power levels $Q$ to be independently selected.

Let the set of available noise-power levels be
\begin{equation}
\mathcal{P}={P_0,P_1,\ldots,P_{Q-1}}.
\label{eq:power_set}
\end{equation}

For a given state-input pair $(\mathbf{s}_t,u_t)$, the mapping $f(\cdot)$ selects one element of $\mathcal{P}$. A convenient deterministic implementation is to assign an index to each state-input pair and map it to a power-level index. For example, if the symbols in $\mathbf{s}_t$ and $u_t$ are interpreted as base-$N$ digits, the transition index can be written as
\begin{equation}
\ell(\mathbf{s}_t,u_t)=u_t+
\sum_{m=1}^{M}u_{t-m}N^m,
\label{eq:transition_index}
\end{equation}
where
\begin{equation}
\ell(\mathbf{s}_t,u_t)\in
{0,1,\ldots,N^{M+1}-1}.
\end{equation}

The corresponding power-level index can then be defined as
\begin{equation}
q(\mathbf{s}_t,u_t)=\ell(\mathbf{s}_t,u_t)\bmod Q,
\label{eq:power_index}
\end{equation}
and the transmitted power is
\begin{equation}
P_t=P_{q(\mathbf{s}_t,u_t)}.
\label{eq:transition_power}
\end{equation}

This definition provides an explicit and reproducible mapping between the trellis transitions and the available noise-power levels. When $Q=N^{M+1}$, the mapping is one-to-one. For smaller values of $Q$, multiple transitions use the same power level while the state-dependent trellis structure is preserved.

For a selected power level $P_t$, the $T$ noise samples corresponding to the $t$th transmitted symbol are generated according to
\begin{equation}
z_{n,t}\sim\mathcal{N}(0,P_t),
\qquad n=1,\ldots,T,
\label{eq:tnm_signal}
\end{equation}
where $T$ denotes the number of transmitted noise samples per symbol.

The available power levels can be selected according to different design criteria. In general, the levels may be uniformly distributed over a prescribed power range, or they may be selected according to a statistical criterion that accounts for the distinguishability of the corresponding received noise distributions. Uniform spacing can be obtained as
\begin{equation}
P_i=P_{\min}+\frac{i}{Q-1}(P_{\max}-P_{\min}),
\qquad i=0,\ldots,Q-1.
\label{eq:uniform_power_levels}
\end{equation}

Alternatively, a statistically motivated design can be obtained using the Bhattacharyya distance between the probability distributions associated with adjacent power levels \cite{fukunaga}, \cite{cutoff_bhattacharyya}. In this work, the latter criterion is considered as one of the possible power-level selection methods and is subsequently adopted for the main simulation experiments after comparing different power-level configurations.

Let $\theta_i$ and $\theta_j$ denote the total received variances corresponding to two power levels. For two zero-mean Gaussian distributions, the Bhattacharyya distance is given by
\begin{equation}
B(\theta_i,\theta_j)=\ln\left(
\frac{\theta_i+\theta_j}
{2\sqrt{\theta_i\theta_j}}
\right).
\label{eq:bhattacharyya}
\end{equation}

A larger value of $B(\theta_i,\theta_j)$ indicates greater statistical separability between the two corresponding distributions. Therefore, the power levels can be designed such that adjacent levels have a prescribed or approximately equal Bhattacharyya distance. Since the receiver noise contributes to the total variance, we define
\begin{equation}
\theta_i=P_i+\sigma_w^2,
\label{eq:theta_definition}
\end{equation}
where $\sigma_w^2$ denotes the receiver noise variance. By imposing a constant Bhattacharyya distance between consecutive levels,
\begin{equation}
B(\theta_i,\theta_{i+1})=B_0,
\label{eq:constant_bhattacharyya}
\end{equation}
the ratio between two consecutive total variance levels becomes constant:
\begin{equation}
\theta_{i+1}=r\theta_i,
\label{eq:variance_geometric}
\end{equation}
where
\begin{equation}
r=
\left(
e^{B_0}+
\sqrt{e^{2B_0}-1}
\right)^2.
\label{eq:variance_ratio}
\end{equation}

Consequently, the corresponding modulation-noise power levels are obtained from
\begin{equation}
P_i=\theta_i-\sigma_w^2.
\label{eq:power_from_variance}
\end{equation}

This approach provides a systematic way of distributing the power levels according to the statistical separability of the received noise distributions rather than their absolute power difference.

At the receiver, let $y_{n,t}$ denote the $n$th received sample corresponding to the $t$th transmitted symbol. The received energy is calculated as
\begin{equation}
E_t=\sum_{n=1}^{T}y_{n,t}^{2}.
\label{eq:received_energy}
\end{equation}

Since the received samples contain both the modulated noise and the receiver noise, the total variance corresponding to a transition with power $P_t$ is
\begin{equation}
P_t'=P_t+\sigma_w^2.
\label{eq:received_variance}
\end{equation}

Under the Gaussian noise model, the negative log-likelihood of the received samples, up to terms independent of the trellis transition, leads to the soft-decision branch metric
\begin{equation}
\gamma(\mathbf{s}_t\rightarrow\mathbf{s}_{t+1})=\frac{T}{2}\ln P_t'
+
\frac{E_t}{2P_t'}.
\label{eq:soft_branch_metric}
\end{equation}

The Viterbi algorithm minimizes the accumulated branch metric over the trellis and therefore estimates the most likely sequence of states and transmitted symbols.

For the hard-decision receiver, the received energy is first quantized into one of the $Q$ available power levels using $Q-1$ decision thresholds. Let the ordered power levels satisfy
\begin{equation}
P_0<P_1<\cdots<P_{Q-1},
\end{equation}
and define the corresponding received variance levels as
\begin{equation}
P_i'=P_i+\sigma_w^2,
\qquad i=0,\ldots,Q-1.
\end{equation}

For two adjacent received variance levels $P_i'$ and $P_{i+1}'$, the ML threshold is obtained by equating their corresponding likelihood functions. The resulting threshold is
\begin{equation}
\mathrm{Th}_{i}^{\mathrm{ML}}=T
\frac{P_i'P_{i+1}'}
{P_{i+1}'-P_i'}
\ln\left(
\frac{P_{i+1}'}{P_i'}
\right),
\qquad i=0,\ldots,Q-2.
\label{eq:nary_ml_threshold}
\end{equation}

The resulting quantized power-level decisions are subsequently processed by the Viterbi decoder. In contrast, the soft-decision receiver avoids this intermediate quantization and directly incorporates the continuous-valued received energy into the branch metric in \eqref{eq:soft_branch_metric}. Consequently, the soft receiver preserves the reliability information contained in the received energy, whereas the hard receiver loses part of this information through quantization.

The three parameters $N$, $M$, and $Q$ play distinct roles in the proposed framework. The modulation order $N$ determines the number of possible input symbols and hence the number of input bits per symbol, $\log_2N$. The memory order $M$ determines the number of trellis states according to $N^M$, while the total number of state-input transitions is $N^{M+1}$. The parameter $Q$ specifies the number of distinct noise-power levels used to represent these transitions. Thus, $Q$ can be selected independently of $N$ and $M$, subject to $Q\leq N^{M+1}$.

In the simulation study, $M=2$ is used unless otherwise specified, while the modulation order is varied among $N=2$, $4$, and $8$. Consequently, the corresponding numbers of trellis states are $4$, $16$, and $64$, and the numbers of trellis transitions are $8$, $64$, and $512$, respectively. The number of available noise-power levels is specified separately by $Q$ and is kept independent of the modulation order when required for a controlled comparison of the different trellis configurations.

Several power-level configurations are initially compared to examine their effect on detection performance. Based on this preliminary comparison, the Bhattacharyya-distance-based power-level design is selected for the subsequent experiments, including the evaluations of BER versus $E_b/N_0$, BER versus the number of samples per symbol, and the comparison of different modulation orders.

\section{Simulation Results}
\label{sec: Simulation}

This section presents simulation results to evaluate the performance of the proposed trellis-based noise modulation scheme and to compare the hard- and soft-decision receiver architectures. In particular, the performance of the conventional hard-demodulation-based Viterbi receiver (RX1) is compared with that of the proposed joint soft Viterbi receiver (RX2). The effects of the noise-power-level selection, traceback depth, $E_b/N_0$, and symbol duration on the bit error rate (BER) are investigated. Unless otherwise stated, the simulations are performed with $M=2$, $N_{\mathrm{bits}}=10{,}000$, while the power levels are selected according to the Bhattacharyya-distance-based design described in the following.

\subsection{Power and $E_b/N_0$ Definitions}

Since information is conveyed through the variance of the transmitted noise waveform, the instantaneous transmit power during symbol interval $t$ is equal to the variance of the transmitted noise samples. Specifically, for
\begin{equation}
z_{t,n}\sim\mathcal{N}(0,P_t),
\end{equation}
the average power of each transmitted noise sample is
\begin{equation}
\mathbb{E}[z_{t,n}^2]=P_t.
\end{equation}
Accordingly, the average transmitted noise power over a block of $N_b$ information symbols is defined as
\begin{equation}
\bar{P}_{\mathrm{tx}}=\frac{1}{N_b} \sum_{t=1}^{N_b}P_t.
\label{eq:average_power}
\end{equation}

Since each information symbol is represented by $T$ independent noise samples, the average energy associated with one transmitted symbol is
\begin{equation}
E_s=T\bar{P}_{\mathrm{tx}}.
\label{eq:Es_definition}
\end{equation}

For the binary scheme, each symbol conveys one information bit. Therefore, the average energy per bit is
\begin{equation}
E_b=T\bar{P}_{\mathrm{tx}}.
\label{eq:Eb_binary}
\end{equation}

Hence, the energy-per-bit-to-noise-power-spectral-density ratio is defined as
\begin{equation}
\frac{E_b}{N_0}=\frac{T\bar{P}_{\mathrm{tx}}} {\sigma_w^2}.
\label{eq:EbN0_binary}
\end{equation}

For the proposed $N$-ary scheme, each symbol conveys
\begin{equation}
b=\log_2(N)
\end{equation}
information bits. Consequently, the average energy per information bit becomes
\begin{equation}
E_b=\frac{T\bar{P}_{\mathrm{tx}}}
{\log_2(N)},
\label{eq:Eb_nary}
\end{equation}
and the corresponding $E_b/N_0$ is given by
\begin{equation}
\frac{E_b}{N_0}=\frac{T\bar{P}_{\mathrm{tx}}}
{\sigma_w^2\log_2(N)}.
\label{eq:EbN0_nary}
\end{equation}

Accordingly, for a desired value of $E_b/N_0$, the channel noise variance is adjusted as
\begin{equation}
\sigma_w^2=\frac{T\bar{P}_{\mathrm{tx}}}
{\left(E_b/N_0\right)\log_2(N)}
\end{equation}
for the $N$-ary scheme, which reduces to
\begin{equation}
\sigma_w^2=\frac{T\bar{P}_{\mathrm{tx}}}
{E_b/N_0}
\end{equation}
for the binary case.

Therefore, all BER results in this work are reported as functions of $E_b/N_0$. For comparisons involving different modulation orders, the noise variance is adjusted according to the above definitions such that the same $E_b/N_0$ is maintained across the considered modulation orders. This provides a normalized comparison that accounts for the different numbers of information bits conveyed by each $N$-ary symbol.

For a given simulation scenario, all receiver architectures use the same transmitted power sequence and channel noise variance. Therefore, the observed performance differences are solely attributable to the employed detection and decoding strategies.

\begin{table}[t]
\centering
\caption{Normalized noise-power levels with a common average transmitted power of $\bar{P}_{\mathrm{tx}}=1.5$.}
\label{tab:power_sets}
\begin{tabular}{c|cccccccc}
\hline
Set & $P_1$ & $P_2$ & $P_3$ & $P_4$ & $P_5$ & $P_6$ & $P_7$ & $P_8$ \\
\hline
$P_1$ & 0.19 & 0.37 & 0.66 & 0.94 &
1.41 & 1.87 & 2.81 & 3.75\\

$P_2$ & 0.55 & 0.82 & 1.09 & 1.36 &
1.64 & 1.91 & 2.18 & 2.45 \\

$P_3$ & 0.15 & 0.54 & 0.92 & 1.31 &
1.69 & 2.08 & 2.46 & 2.85 \\

$P_4$ & 0.13 & 0.52 & 0.91 & 1.30 &
1.69 & 2.09 & 2.48 & 2.87\\

$P_B$ & 0.20 & 0.36 & 0.57 & 0.89 &
1.32 & 1.93 & 2.77 & 3.96\\
\hline
\end{tabular}
\end{table}

\begin{table*}[!t]
\caption{BER comparison of different noise-power-level designs at
$E_b/N_0=10$ dB.}
\label{tab:power_design_comparison}
\centering
\renewcommand{\arraystretch}{1.15}
\begin{tabular}{lcccc}
\hline
\textbf{Power-level design} &
\textbf{Hard Viterbi (Midpoint)} &
\textbf{Hard Viterbi (ML threshold)} &
\textbf{Soft Viterbi} &
\textbf{Symbol-by-Symbol ML} \\
\hline

$\mathcal{P}_{\mathrm{B}}$ &
0.2106 & 0.2074 & \textbf{0.0194} & 0.4314 \\

$\mathcal{P}_1$ &
0.2176 & 0.2154 & \textbf{0.0236} & 0.4248 \\

$\mathcal{P}_2$ &
0.3176 & 0.3206 & \textbf{0.1591} & 0.4319 \\

$\mathcal{P}_3$ &
0.2481 & 0.2412 & \textbf{0.0741} & 0.3868 \\

$\mathcal{P}_4$ &
0.2382 & 0.2435 & \textbf{0.0717} & 0.3818 \\

\hline
\end{tabular}
\end{table*}

\subsection{Binary Noise Modulation}
We first evaluate the performance of the proposed hard- and soft-decision Viterbi receivers in the binary noise modulation scheme. The effects of the traceback depth, $E_b/N_0$, and the number of noise samples per symbol on the BER performance are investigated.

\paragraph{Selection of Noise-Power Levels}

To investigate the effect of the noise-power level configuration, five different designs were considered, including four representative configurations, denoted by $\mathcal{P}_1$--$\mathcal{P}_4$, and a Bhattacharyya-based design, $\mathcal{P}_{\mathrm{B}}$. The corresponding power levels are listed in Table~\ref{tab:power_sets}. All configurations were evaluated under identical conditions with $M=2$, $T=20$, $N_{\mathrm{bits}}=10{,}000$, $E_b/N_0=10$~dB, and $D=10$.

The corresponding BER results are summarized in Table~\ref{tab:power_design_comparison}. The results show that the power-level configuration significantly affects the detection performance, as insufficient separation between adjacent levels increases detection ambiguity. The soft-decision Viterbi receiver consistently outperforms the hard-decision receivers for all considered designs. The Bhattacharyya-based configuration is therefore adopted as the baseline power-level design in the subsequent simulations.

\begin{figure}[!t]
\centering
{\includegraphics[width=.9\linewidth]{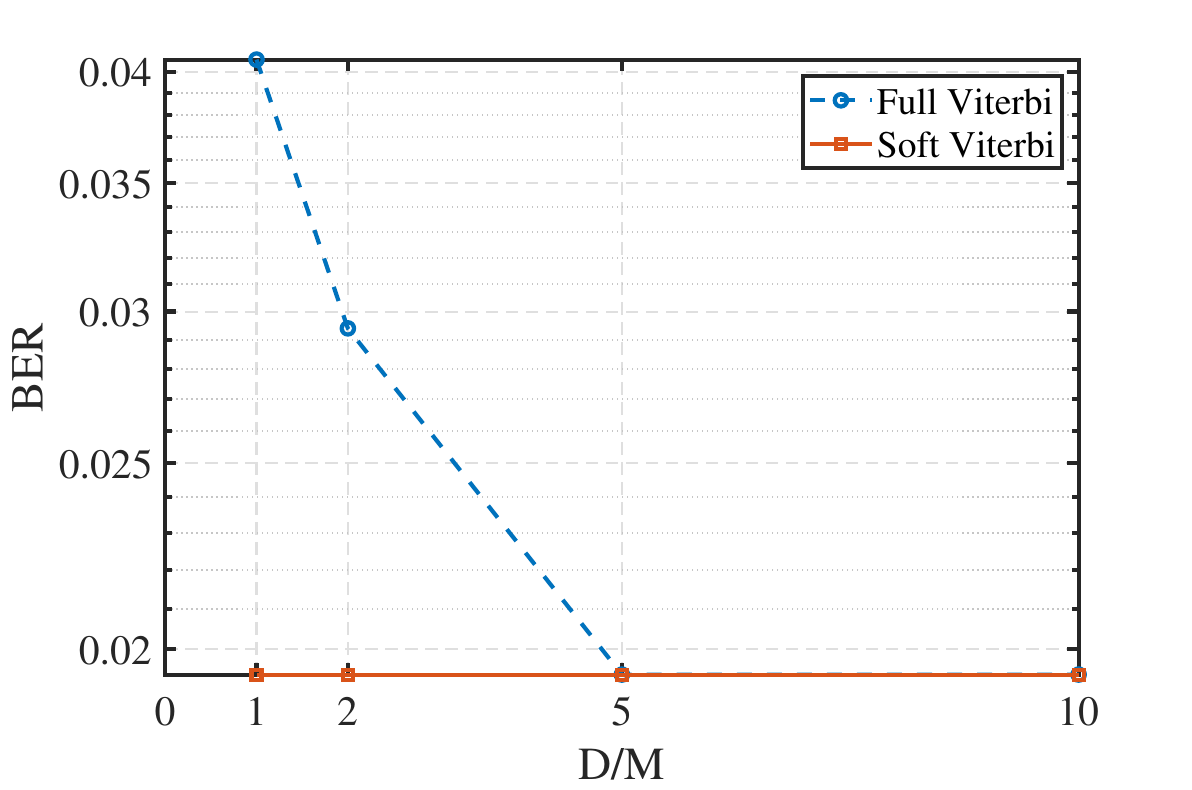}}%
\caption{BER performance versus traceback depth for the soft Viterbi receiver.}
\label{fig:ber_traceback}
\end{figure}

\paragraph{Effect of Traceback Depth}
The effect of the traceback depth on the performance of the proposed soft Viterbi receiver was investigated using the Bhattacharyya-based power-level configuration, and for $E_b/N_0=10~dB$. As shown in Fig.~\ref{fig:ber_traceback}, increasing the traceback depth significantly improves the BER performance for small values of $D$. However, the performance approaches that of the full-traceback Viterbi receiver as the traceback depth increases, and no further improvement is observed beyond $D=5M$. Therefore, $D=5M$ provides a suitable trade-off between decoding performance and traceback delay and is adopted for the subsequent simulations.

\begin{figure}[!t]
\centering
{\includegraphics[width=.9\linewidth]{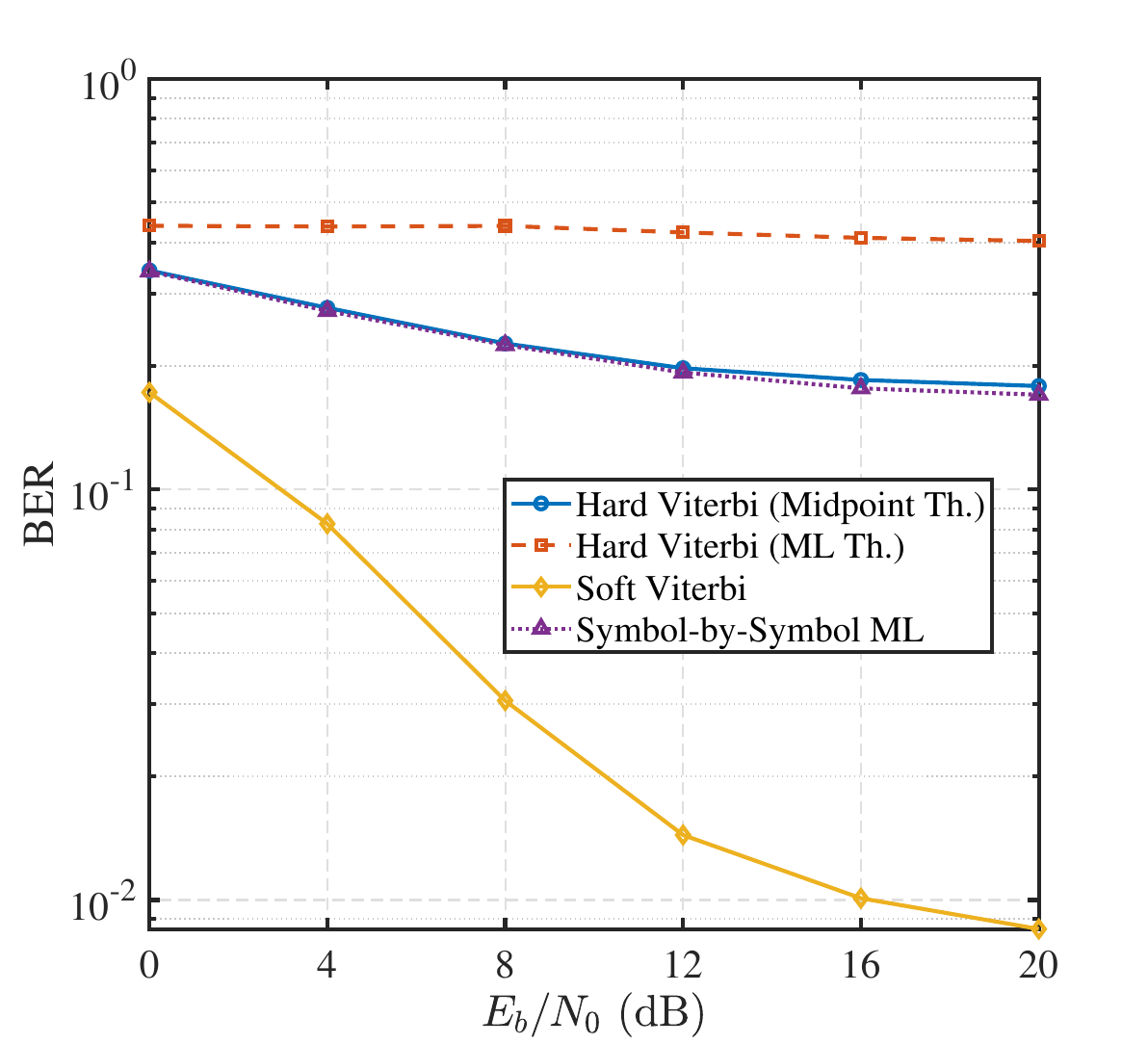}}%
\caption{BER performance versus $E_b/N_0$ for the hard and soft Viterbi receivers.}
\label{fig:ber_snr}
\end{figure}

\paragraph{BER Performance versus $E_b/N_0$}

The BER performance of the binary filtering-based receivers was evaluated over a range of $E_b/N_0$ values using the Bhattacharyya-based power-level configuration. As shown in Fig.~\ref{fig:ber_snr}, the BER of all receivers increases as the $E_b/N_0$ decreases. The soft receiver consistently achieves the best performance, while the two hard-decision receivers with midpoint and ML-based thresholds exhibit very similar performance.

The performance advantage of the soft receiver results from directly exploiting the continuous-valued energy observations, whereas hard detection quantizes these observations and discards reliability information. The small gap between the two hard thresholds further indicates that the main performance loss is caused by hard quantization rather than the particular threshold choice.

For the binary rate-one filtering configuration, the filtering operation is invertible and introduces no coding redundancy. Therefore, the gain of the soft receiver should not be interpreted as a conventional coding gain, but rather as the benefit of soft energy-based detection over hard quantization. The state-dependent $N$-ary scheme considered subsequently provides the more general trellis-based structure of the proposed framework.

\begin{figure}[!t]
\centering
{\includegraphics[width=.9\linewidth]{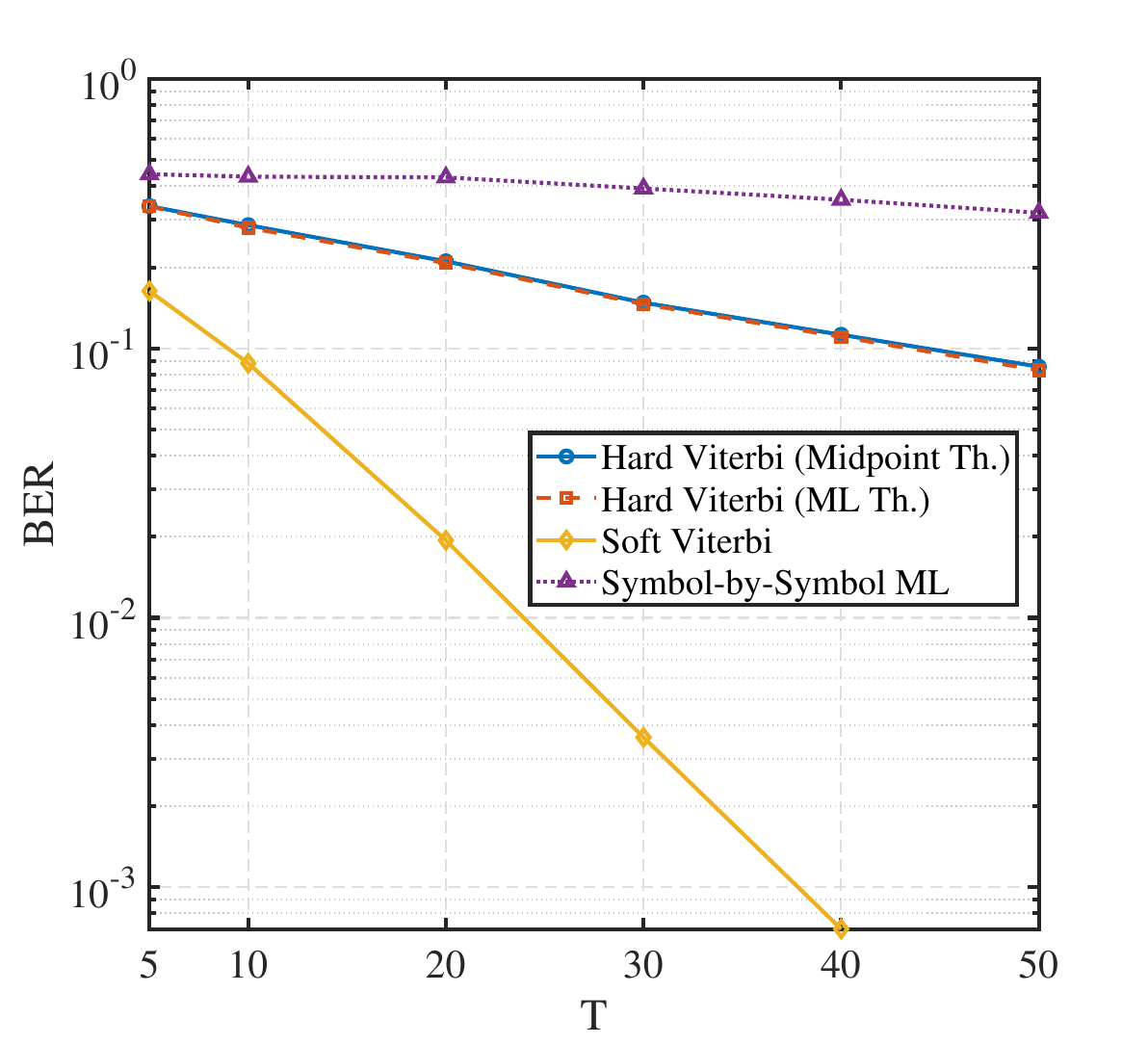}}%
\caption{BER performance versus symbol duration $T$ for the hard and soft Viterbi receivers.}
\label{fig:ber_T}
\end{figure}

\paragraph{Effect of Symbol Duration}

The effect of the symbol duration $T$ on the binary filtering-based receivers was also investigated using the Bhattacharyya-based power-level configuration, and for $E_b/N_0=10~dB$. As shown in Fig.~\ref{fig:ber_T}, increasing $T$ improves the performance of all receivers because more noise samples provide a more reliable estimate of the received noise power.

The improvement is most pronounced for the soft receiver, which directly exploits the continuous-valued energy observations, whereas hard detection loses reliability information through energy quantization. The symbol-by-symbol ML detector also exhibits inferior performance because it processes the observations independently without exploiting the filtering relationship across successive symbols.

Since the binary filtering operation is rate one and invertible, the observed improvement should not be interpreted as a conventional coding gain. Rather, these results primarily illustrate the benefit of soft processing and longer energy observations in the binary noise modulation configuration.

\subsection{N-ary Noise Modulation}
To further evaluate the applicability of the proposed receivers, the binary noise modulation scheme is extended to an $N$-ary configuration with multiple noise-power levels. The performance of the hard- and soft-decision Viterbi receivers is then investigated for different modulation orders in terms of BER, $E_b/N_0$, and the number of noise samples per symbol.

\paragraph{BER Performance versus $E_b/N_0$}

\begin{figure}[!t]
\centering
{\includegraphics[width=.9\linewidth]{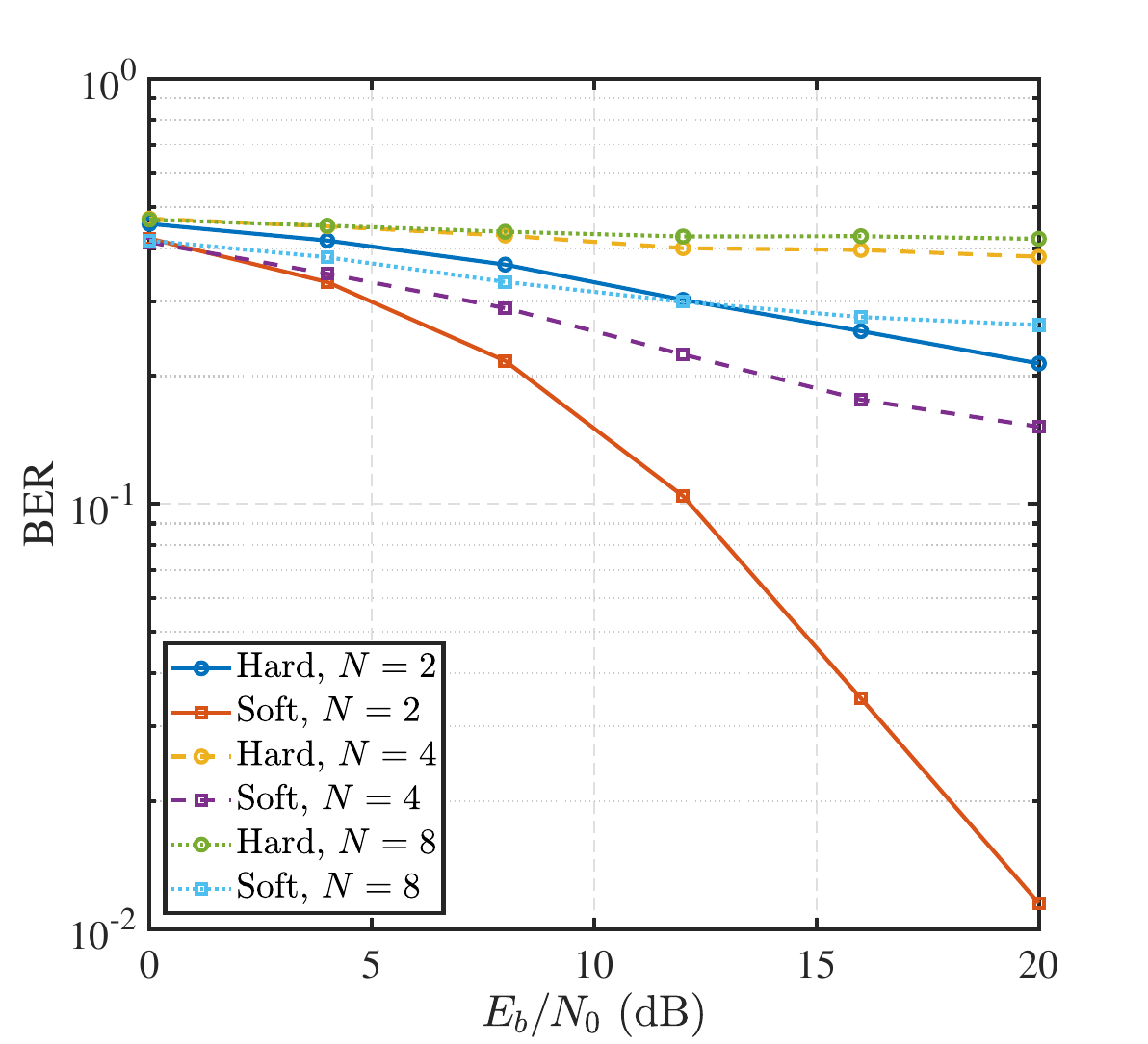}}%
\caption{BER performance of the hard- and soft-decision Viterbi receivers versus $E_b/N_0$.}
\label{fig:ber_EbN0_Nary}
\end{figure}

Fig.~\ref{fig:ber_EbN0_Nary} presents the BER performance of the hard- and soft-decision Viterbi receivers for different modulation orders. The results are obtained for $N=2$, $4$, and $8$, with memory order $M=2$ and $T=20$ Gaussian noise samples per symbol. To ensure a fair comparison among different modulation orders, all configurations are normalized to the same average transmitted noise power, with $\bar{P}_{\mathrm{tx}}=1.5$.

As observed in Fig.~\ref{fig:ber_EbN0_Nary}, the soft-decision Viterbi receiver consistently provides better performance than the hard-decision receiver for all considered modulation orders. This improvement results from directly incorporating the received noise energy into the likelihood-based branch metric, whereas the hard-decision receiver first quantizes the observed energy and consequently discards part of the available statistical information. The performance gap becomes particularly evident as the noise level decreases, demonstrating the advantage of soft energy-based sequence detection.

The BER performance also degrades as the modulation order increases. For the considered trellis structure, the number of states and transitions grows as $N^M$ and $N^{M+1}$, respectively. Since each transition is assigned to a distinct noise-power level in this implementation, increasing $N$ also substantially increases the number of available power levels. With a common average transmitted power and a fixed power range, the power levels therefore become more densely packed as $N$ increases, reducing their statistical separation. This makes reliable discrimination between neighboring transitions more difficult and results in degraded detection performance for higher-order modulation.

These results illustrate the fundamental trade-off of the proposed $N$-ary noise modulation scheme. Increasing the modulation order increases the number of information bits conveyed by each symbol, but simultaneously increases trellis complexity and reduces the separation between the noise-power levels. The soft-decision Viterbi receiver mitigates part of this degradation by exploiting the complete received-energy information, making it particularly beneficial when the number of available power levels becomes large.

\begin{figure}[!t]
\centering
{\includegraphics[width=.9\linewidth]{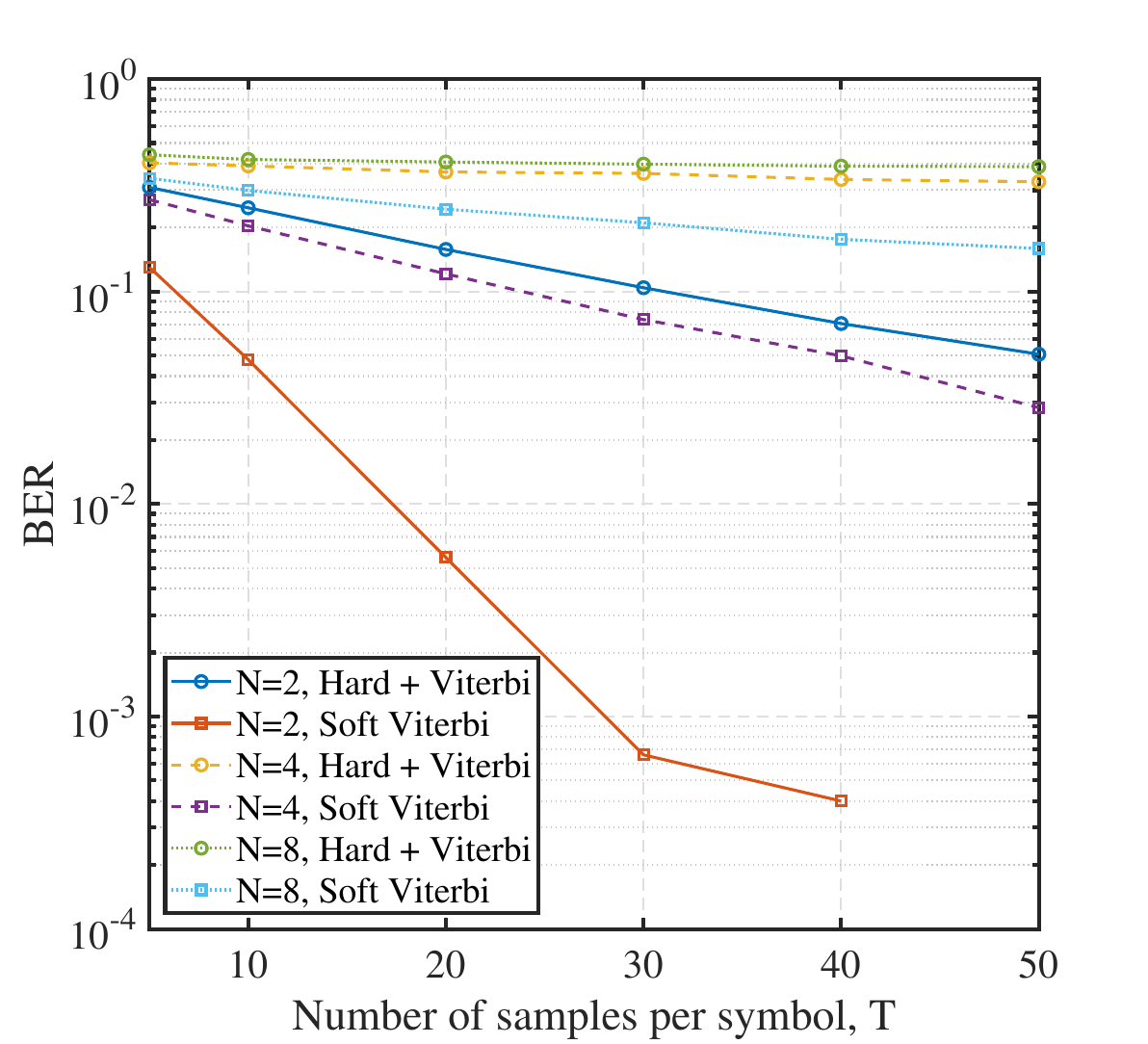}}%
\caption{BER performance versus symbol duration $T$ for the hard and soft Viterbi receivers.}
\label{fig:ber_T_Nary}
\end{figure}

\paragraph{Effect of Symbol Duration}

Fig.~\ref{fig:ber_T_Nary} presents the BER performance as a function of the number of noise samples per symbol, $T$, at a fixed $E_b/N_0$ of $10$~dB. The results are reported for $N=2$, $4$, and $8$, with $M=2$. Increasing $T$ improves the detection performance for both receivers. This improvement is attributed to the increased number of observations available for estimating the received noise power, which reduces the uncertainty of the energy statistic used for detection. In addition, for a fixed per-sample $E_b/N_0$, increasing $T$ increases the accumulated energy over each symbol, further improving the reliability of the energy-based detection. The soft-decision receiver consistently outperforms the hard-decision receiver, as it exploits the received energy directly in the likelihood-based branch metric rather than reducing it to a quantized decision. The performance degradation associated with higher modulation orders is also evident, since the increased number of power levels makes reliable discrimination more challenging. Overall, the results demonstrate that increasing the observation length can substantially improve the reliability of trellis-based noise modulation, at the cost of increased symbol duration and receiver processing.

\subsection{Performance under Fading Channels}
\label{subsec:fading}
To further investigate the wireless applicability of the proposed trellis-based noise modulation scheme, its performance was evaluated over Rayleigh and Rician fading channels. A block-fading model was considered, in which the channel coefficient remains constant during the $T$ samples corresponding to one transmitted symbol and varies independently from one symbol interval to another. The received signal is modeled as
\begin{equation}
r_{t,n}=h_t z_{t,n}+w_{t,n},
\end{equation}
where $h_t$ denotes the fading coefficient, $z_{t,n}$ is the transmitted noise sample, and $w_{t,n}\sim\mathcal{N}(0,\sigma_w^2)$ represents the receiver noise. Consequently, conditioned on the instantaneous channel gain $g_t=|h_t|^2$, the received samples have variance
\begin{equation}
P_{t,\mathrm{eff}}'=g_tP_t+\sigma_w^2.
\label{eq:fading_effective_variance}
\end{equation}
When channel state information (CSI) is available at the receiver, the instantaneous channel gain is incorporated into the soft Viterbi branch metric by replacing $P_t'$ with $P_{t,\mathrm{eff}}'$. In the absence of CSI, the receiver employs the AWGN-based metric without adaptation to the instantaneous fading coefficient.

\begin{figure}[!t]
\centering
{\includegraphics[width=.9\linewidth]{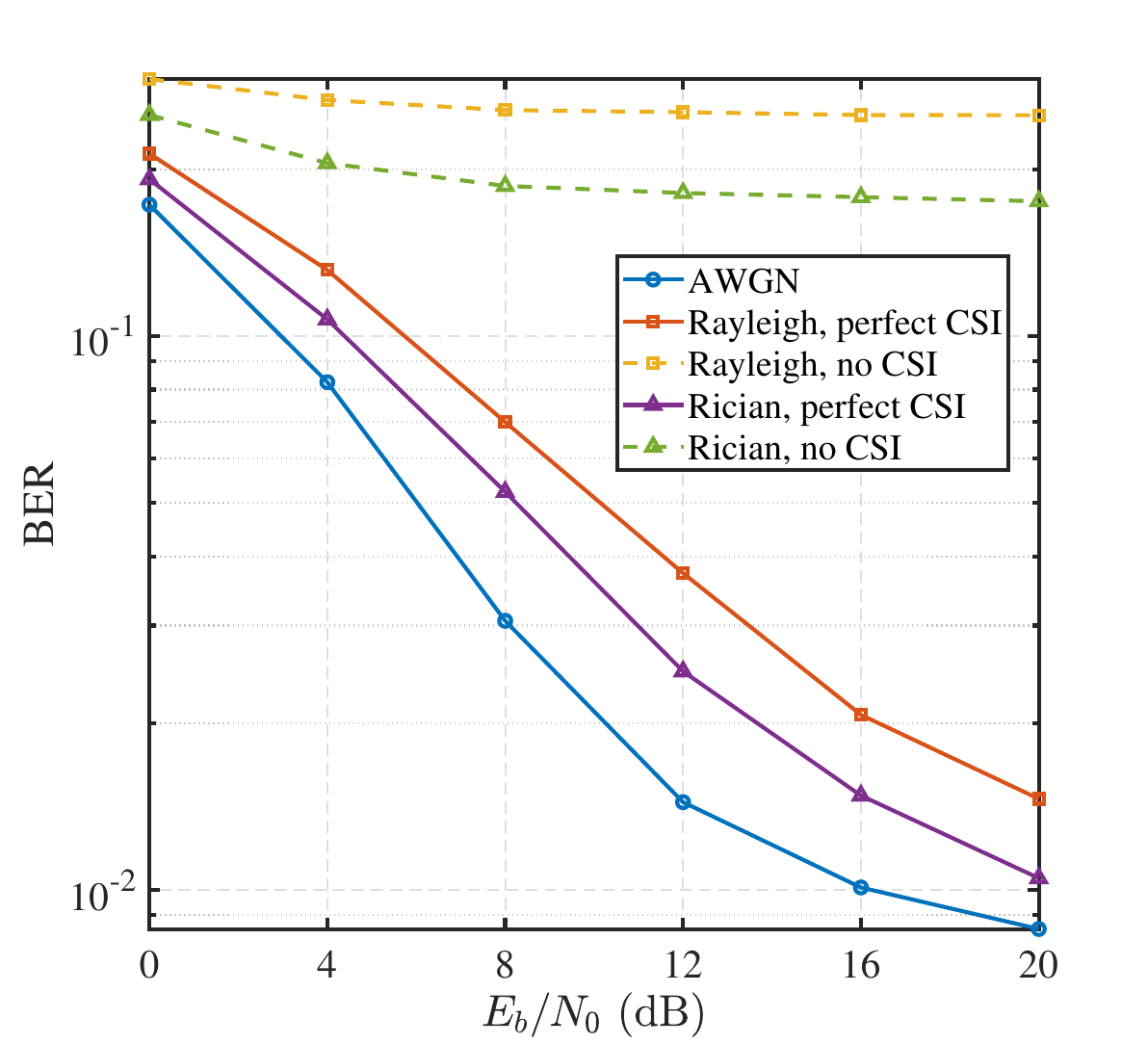}}%
\caption{BER performance of the proposed soft Viterbi receiver over AWGN, Rayleigh, and Rician channels.}
\label{fig:fading_ber}
\end{figure}

Fig.~\ref{fig:fading_ber} compares the BER performance of the proposed soft Viterbi receiver over AWGN, Rayleigh fading, and Rician fading channels. For the fading channels, both perfect CSI and no-CSI receiver configurations are considered. As expected, the AWGN channel provides the best overall performance because no random amplitude fluctuations are introduced. Under fading conditions, the receiver with perfect CSI significantly outperforms the corresponding no-CSI receiver over the entire $E_b/N_0$ range. This improvement results from adapting the likelihood metric to the instantaneous channel gain and, consequently, to the effective received noise variance.

The Rician channel generally achieves better performance than the Rayleigh channel due to the presence of a line-of-sight component, which reduces the severity of deep fading. These results demonstrate that the proposed trellis-based receiver can operate effectively under fading channels when the receiver accounts for the instantaneous channel conditions. In contrast, ignoring the fading coefficient causes a substantial performance degradation, particularly in the Rayleigh channel, where the random channel gain can significantly distort the variance levels used to convey information.

\section{Conclusion}
\label{sec:con}

This paper investigated finite-state and trellis-based detection structures for noise modulation. A binary filtering-based configuration was first considered as an illustrative finite-memory framework for comparing hard- and soft-decision energy-based receivers. Since the binary filtering operation is rate one and invertible, it does not provide conventional coding redundancy; instead, it highlights the benefit of preserving the received energy information in sequence detection.

The main contribution of the paper is the extension to an $N$-ary state-dependent trellis-based noise modulation scheme, in which the transmitted noise-power level is jointly determined by the trellis state and the input symbol. This structure introduces nontrivial transition constraints that can be exploited by the Viterbi algorithm for sequence-based detection. A likelihood-based soft Viterbi receiver was developed to directly incorporate the received energy into the branch metric, avoiding the information loss associated with intermediate hard decisions.

Simulation results demonstrated that the soft Viterbi receiver consistently outperforms the hard-decision and symbol-by-symbol receivers. The results also showed that the observation length, modulation order, and noise-power-level configuration significantly affect the detection performance. In particular, the Bhattacharyya-distance-based power-level design provides a systematic approach for selecting statistically distinguishable noise-power levels. Finally, the proposed receiver was evaluated over Rayleigh and Rician fading channels, where incorporating channel state information into the detection metric substantially improves performance.

Overall, the results demonstrate the potential of trellis-based soft-decision sequence detection as a framework for improving the reliability of binary and higher-order noise modulation systems.

\end{document}